\documentclass[aps,prd,eqsecnum,preprint,tightenlines,nofootinbib]{revtex4-2}
\usepackage{graphicx,latexsym}

\def\bea{\begin{eqnarray}}
\def\eea{\end{eqnarray}}
\def\be{\begin{equation}}
\def\ee{\end{equation}}

\newcommand{\Pminus}{{\cal P}^-}
\newcommand{\Pfree}{\Pminus_0}
\newcommand{\Pint}{\Pminus_{\rm int}}
\newcommand{\Pvac}{P_{\rm vac}^-}
\newcommand{\vac}{|{\rm vac}\rangle}
\newcommand{\deps}{\delta_\epsilon}

\begin{document}

\title{Tadpoles and vacuum bubbles in light-front quantization
}
\author{S.S.~Chabysheva}
\affiliation{Department of Physics, University of Idaho, Moscow ID 83844 USA}
\author{J.R.~Hiller}
\affiliation{Department of Physics, University of Idaho, Moscow ID 83844 USA}
\affiliation{Department of Physics and Astronomy,
University of Minnesota-Duluth,
Duluth, Minnesota 55812 USA}

\date{\today}

\begin{abstract}

We develop a method by which vacuum transitions may be included
in light-front calculations.  This allows tadpole contributions
which are important for symmetry-breaking effects and yet
are missing from standard light-front calculations.  These
transitions also dictate a nontrivial vacuum and contributions
from vacuum bubbles to physical states.  In nonperturbative
calculations these separate classes of contributions
(tadpoles and bubbles) cannot be filtered; instead,
we regulate the bubbles and subtract the vacuum energy 
from the eigenenergy of physical states.  The key is
replacement of momentum-conserving delta functions with
model functions of finite width; the width becomes the
regulator and is removed after subtractions.  The 
approach is illustrated in free scalar theory, quenched
scalar Yukawa theory, and in a limited Fock-space truncation
of $\phi^4$ theory.

\end{abstract}


\maketitle

\section{Introduction}
\label{sec:Introduction}

\begin{figure}[hb]
\vspace{0.2in}
\begin{tabular}{cc}
\includegraphics[width=5cm]{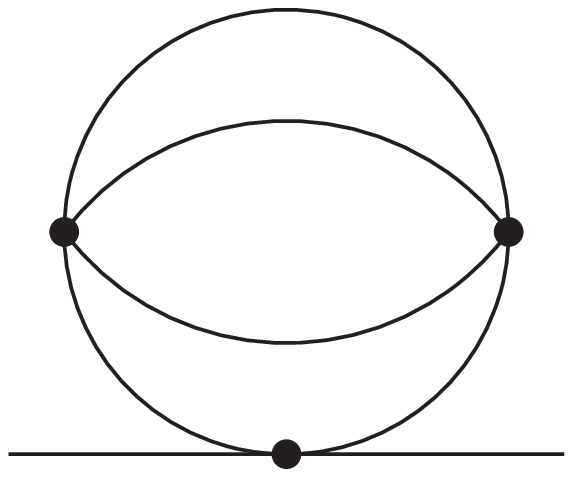} &
\includegraphics[width=5cm]{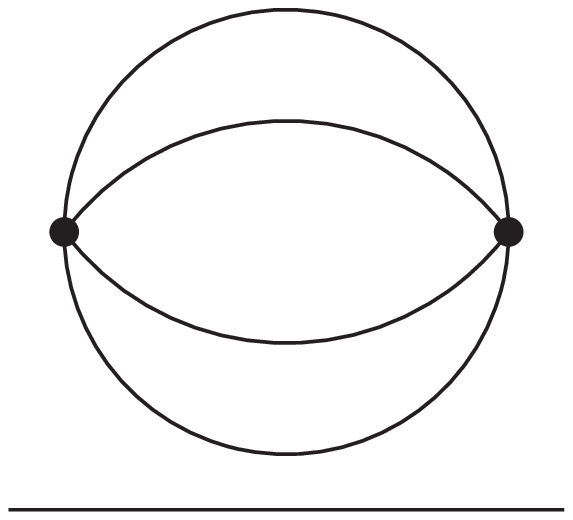} \\
(a) & (b)
\end{tabular}
\caption{\label{fig:tadpole}
Tadpole graph (a) and vacuum bubble (b) in $\phi^4$ theory.  Note the 
momentum-conserving transitions to and from
the vacuum that imply light-front zero-mode contributions.
}
\end{figure}

Recent calculations of the critical coupling in two-dimensional $\phi^4$ 
theory~\cite{RychkovVitale,Rychkov,Pelissetto,DePalma,BCH,Katz1,Katz2,%
Katz3,Katz4,Serone,Kadoh,Heymans,VaryNewDLCQ}
have shown that there is a discrepancy in the nonperturbative equivalence of
equal-time and light-front quantization.  Although this discrepancy can be
explained with a computed shift in the renormalized 
mass~\cite{SineGordon,BCH,Katz1,Katz3}, this explanation
is a correction to the light-front calculation, rather than a direct calculation.
In addition, calculations with coordinates that interpolate between equal-time
and light-front quantizations~\cite{Hornbostel,Ji}, indicate that the light-front limit
should obtain the same critical coupling as obtained in
equal time quantization~\cite{ETtoLF}.  The key is the inclusion of tadpole 
contributions, which on the light front requires zero modes, modes
with zero longitudinal momentum, to represent transitions to and from
the vacuum, as illustrated in Fig.~\ref{fig:tadpole}(a).\footnote{These figures
were drawn with JaxoDraw~\protect\cite{JaxoDraw}.}  In a nonperturbative 
calculation, where one cannot pick and choose classes of diagrams, the
presence of vacuum transitions necessarily imports (divergent) vacuum 
bubbles, of a sort shown in Fig.~\ref{fig:tadpole}(b),
as well as tadpoles.\footnote{This is a separate question
from perturbative equivalence, which has been generally established.
For recent discussions, see \protect\cite{MannheimPLB,Mannheim,Polyzou}.}

Zero-mode contributions to physical states and the corresponding
nontriviality of the light-front vacuum~\cite{Collins,Martinovic,MannheimPLB,Mannheim} are 
of broader interest than just the critical coupling in $\phi^4$ theory.  They 
enter into any discussion of symmetry breaking, such as the Higgs
mechanism, and of vacuum condensates~\cite{ChiralSymmetry}.  More recently
they have been identified as possible contributions
to higher-twist distribution functions~\cite{GPDs}.  For these reasons,
we explore a possible method for inclusion, within the
context of two-dimensional scalar theories; extension to
three and four dimensions should be straightforward.\footnote{For
alternative methods, see \protect\cite{McCartor} and \protect\cite{Herrmann}.}

Contributions such as tadpoles and vacuum bubbles that
involve transitions to and from the vacuum must rely on
terms normally excluded from light-front Hamiltonians.
These are terms with only creation operators or only
annihilation operators.  With light-front longitudinal
momenta constrained to be nonnegative, momentum conservation
requires that the operators create or annihilate
zero momentum.  On this basis they are always dropped.
However, depending on the zero-momentum behavior of the 
Fock-space wave functions, matrix elements of such terms
need not be zero.

For example, consider light-front quantization~\cite{Dirac,BPP,Carbonell,Miller,Heinzl,Burkardt,Hiller}
of a two-dimensional scalar theory.  We define light-front coordinates~\cite{Dirac}
and momenta as $x^\pm=t\pm z$ and $p^\pm=E\pm p_z$, with $x^+$ chosen as the 
light-front time.  The mass-shell condition 
for the total two-momentum $(P^-,P^+)$ is then $M^2=P^+P^-$,
which gives the fundamental bound-state eigenproblem as
\be
\Pminus|\psi(P^+)\rangle=\frac{M^2}{P^+}|\psi(P^+)\rangle,
\ee
where $\Pminus$ is the light-front Hamiltonian.
A typical Fock-state wave function for an eigenstate of this Hamiltonian 
satisfies an equation of the following form:
\be
\sum_i^n \frac{\mu^2}{p_i^+}\psi_n(p_1^+,\ldots,p_n^+)
+\frac{1}{\sqrt{p_1^+p_2^+p_3^+\ldots}}\times\mbox{contributions from other Fock sectors}=\frac{M^2}{P^+}\psi_n,
\ee
with $n$ the number of constituents.
The symmetry for bosons then requires that the small momentum
behavior of this wave function is
\be  \label{eq:behavior}
\psi_n\sim \frac{1}{\sqrt{\prod_i^n p_i^+}\sum_i^n\frac{1}{p_i^+}}.
\ee
The matrix element of a vacuum transition in $\phi^4$ theory reduces to
\bea
\lefteqn{\langle \psi_{n+4}|\int \frac{\prod_i^4 dp_i^+}{\sqrt{\prod_i^4 p_i^+}}
  \delta(\sum_i^4 p_i^+)\prod_i^4 a^\dagger(p_i^+)|\psi_n\rangle}&& \\
&&\sim \int \frac{\prod_i^{n+4} dp_i^+}{\prod_i^{n+4} p_i^+}
           \frac{\delta(\sum_i^4 p_{n+i}^+)}
              {\left(\sum_i^{n+4}\frac{1}{p_i^+}\right)\left(\sum_i^n\frac{1}{p_i^+}\right)}
              \delta(P^+-\sum_i^n p_i^+).
\eea
With $Q\equiv\sum_i^4 p_{n+i}^+$, $p_{n+i}^+=x_iQ$, and 
$\prod_{i=n+1}^{n+4} dp_i^+=Q^3dQ\prod_i^4 dx_i\delta(1-\sum_i^4x_i)$, this becomes
\bea
\lefteqn{\langle \psi_{n+4}|\int \frac{\prod_i^4 dp_i^+}{\sqrt{\prod_i^4 p_i^+}}
  \delta(\sum_i^4 p_i^+)\prod_i^4 a^\dagger(p_i^+)|\psi_n\rangle} & & \\
&& \sim \int dQ \delta(Q)  \int \frac{\prod_i^4 dx_i \delta(1-\sum_i^4 x_i)}
                                      {\left(\prod_i^4 x_i\right)\left(\sum_i^4\frac{1}{x_i}\right)}
   \int \frac{\prod_i^n dp_i^+ \delta(P^+-\sum_i^n p_i^+)}
      {\left(\prod_i^n p_i^+\right)\left(\sum_i^n\frac{1}{p_i^+}\right)},
   \nonumber
\eea
which is finite and nonzero.  Thus, such vacuum-transition terms cannot be ignored automatically.

Vacuum transitions also generate vacuum bubbles which make contributions proportional
to the size $L$ of the spatial dimension as expressed through 
$4\pi\delta(0)=\int dx^-\equiv L$.  A nonperturbative calculation
requires a cutoff, to regulate this infinity, and a subtraction of
the vacuum energy from any eigenenergy of a physical state.  We
regulate by replacing delta functions of momentum with model functions $\deps$
that have a width parameter $\epsilon$ and take the limit of 
$\epsilon\rightarrow0$ at the end of a calculation.\footnote{This is
equivalent to the regulator used in \protect\cite{MannheimPLB} for the
calculation of a time-ordered product.  The parameter $R$ in their
Eq.~(22), introduced as the radius of a circle approaching infinity,
is essentially $1/\epsilon$.  However, the circle at infinity is 
something that arises in diagrammatic calculations rather than 
nonperturbative eigenvalue problems.}
For the (nontrivial) vacuum state, we compute a finite energy density,
with the model parameter $\epsilon$ related to the
spatial volume $L$ in a model-dependent way: $4\pi\deps(0)=L$.

For any finite width $\epsilon$, there will be additional modes
present in any calculation.  We call these ephemeral modes, since
they are not zero modes but instead disappear in the limit of
zero width.  The remaining imprint is essentially a zero-mode
contribution, but obtained as a limit.  Contributions to massive
states, beyond the vacuum-energy shift and tadpoles, are generally 
negligible for weak coupling; however,
for strong coupling, the Fock-state momentum wave functions
can be become broad enough that they overlap with ephemeral
modes.  Depending on the zero-momentum behavior of these
wave functions, there can be additional contributions from 
vacuum transitions.

Such contributions cannot be readily captured with the DLCQ
formalism~\cite{PauliBrodsky}, and DLCQ calculations with constrained
zero modes~\cite{Robertson,Pinsky,CH} are incomplete.  For good resolution of the
ephemeral modes, the DLCQ resolution $K$ must satisfy $1/K\ll \epsilon/P^+$.
Also, the integrals that must be represented by the 
rectangular DLCQ grid are highly singular, for which the
grid is ill suited.  Calculations would be best undertaken
with a basis function expansion, for which matrix elements can
be computed once and for all with an adaptive Monte Carlo integration, such
as is available in the {\em VEGAS} package~\cite{vegas}.

Even with antiperiodic boundary conditions, the DLCQ approach
cannot neglect zero modes.  With such boundary conditions
one can avoid the constraint equation, but
the approximation to the integral
operators in $\Pminus$ is the midpoint rule with an error
no better than the $1/K^2$ for periodic boundary 
conditions, where the integrals are approximated by
the trapezoidal rule.\footnote{Without a solution to the 
constraint equation, the error in DLCQ with periodic 
boundary conditions is of order $1/K$, unless the endpoints
(the zero modes) make no contribution to the integrals.}
The trouble is that the coefficient of the $1/K^2$ correction
is small only if the integrand is slowly varying.  If instead
there is rapid variation, such as can happen near zero momentum,
the approximation becomes quite poor except at very high
resolution. In other words, if zero-mode contributions are
important, antiperiodic boundary conditions do not provide
an approximation any better than periodic boundary conditions.

To explore the inclusion of vacuum transitions, we first consider $\phi^4$ theory in
more detail; in Sec.~\ref{sec:phi4} we consider the leading
tadpole and vacuum-bubble contributions.  Next, in Sec.~\ref{sec:shifted},
we develop an analytic solution for a free scalar as a 
generalized coherent state of ephemeral modes.  The vacuum
bubble contributions replicate the one-loop calculation
emphasized by Collins~\cite{Collins}.  We also consider
the solution for a shifted scalar with nonzero
vacuum expectation value. This is
done in the continuum, without interpolation from
equal-time quantization and without discretization.
Finally, we consider quenched scalar Yukawa theory in lowest-order
Fock truncation in Sec.~\ref{sec:QSY}, to see the subtraction of the vacuum
energy of the neutral scalar in the charge-zero sector
from the dressed scalar energy in the charge-one sector.
Numerical calculations are postponed to future work.

\section{Lowest-order $\phi^4$ theory}  \label{sec:phi4}

The Lagrangian for two-dimensional $\phi^4$ theory is
\be
{\cal L}=\frac12(\partial_\mu\phi)^2-\frac12\mu^2\phi^2-\frac{\lambda}{4!}\phi^4,
\ee
where $\mu$ is the mass of the boson and $\lambda$ is the coupling constant.
The light-front Hamiltonian density is
\be
{\cal H}=\frac12 \mu^2 \phi^2+\frac{\lambda}{4!}\phi^4.
\ee
The mode expansion for the field is\footnote{For convenience 
we drop the $+$ superscript and will from here on 
write light-front momenta such as $p^+$ as just $p$.} 
\be \label{eq:phi}
\phi(x^+=0,x^-)=\int \frac{dp}{\sqrt{4\pi p}}
   \left\{ a(p)e^{-ipx^-/2} + a^\dagger(p)e^{ipx^-/2}\right\}.
\ee
The nonzero commutation relation is
\be \label{eq:acomm}
[a(p),a^\dagger(p')]=\delta(p-p').
\ee

The light-front Hamiltonian is 
$\Pminus=\Pfree+\Pint$,
with
\bea \label{eq:Pminus}
\Pfree&=&\int dp \frac{\mu^2}{p} a^\dagger(p)a(p)
   +\frac{\mu^2}{2}\int \frac{dp_1 dp_2}{\sqrt{p_1p_2}}\deps(p_1+p_2)
                 \left[a(p_1)a(p_2)+a^\dagger(p_1)a^\dagger(p_2)\right],  \\
\Pint&=&\Pminus_{04}+\Pminus_{40}+\Pminus_{22}+\Pminus_{13}+\Pminus_{31},
\eea
where
\bea
\label{eq:Pminus04}
\Pminus_{04}&=&\frac{\lambda}{24}\int \frac{dp_1dp_2dp_3dp_4}
                              {4\pi \sqrt{p_1p_2p_3p_4}} 
     \deps(\sum_i^4 p_i) a(p_1)a(p_2)a(p_3)a(p_4), \\
\label{eq:Pminus40}
\Pminus_{40}&=&\frac{\lambda}{24}\int \frac{dp_1dp_2dp_3dp_4}
                              {4\pi \sqrt{p_1p_2p_3p_4}} 
      \deps(\sum_i^4 p_i)a^\dagger(p_1)a^\dagger(p_2)a^\dagger(p_3)a^\dagger(p_4), \\
\label{eq:Pminus22}
\Pminus_{22}&=&\frac{\lambda}{4}\int\frac{dp_1 dp_2}{4\pi\sqrt{p_1p_2}}
       \int\frac{dp'_1 dp'_2}{\sqrt{p'_1 p'_2}} 
       \delta(p_1 + p_2-p'_1-p'_2) \\
 && \rule{2in}{0mm} \times a^\dagger(p_1) a^\dagger(p_2) a(p'_1) a(p'_2),
   \nonumber \\
\label{eq:Pminus13}
\Pminus_{13}&=&\frac{\lambda}{6}\int \frac{dp_1dp_2dp_3}
                              {4\pi \sqrt{p_1p_2p_3(p_1+p_2+p_3)}} 
     a^\dagger(p_1+p_2+p_3)a(p_1)a(p_2)a(p_3), \\
\label{eq:Pminus31}
\Pminus_{31}&=&\frac{\lambda}{6}\int \frac{dp_1dp_2dp_3}
                              {4\pi \sqrt{p_1p_2p_3(p_1+p_2+p_3)}} 
      a^\dagger(p_1)a^\dagger(p_2)a^\dagger(p_3)a(p_1+p_2+p_3).
\eea
The subscripts indicate the number of creation and annihilation operators
in each term.

To isolate the contribution from the tadpole and vacuum bubble in Fig.~\ref{fig:tadpole},
we consider only two terms in the Fock-state expansion of the eigenstate
\be
|\psi(P)\rangle=\psi_1 a^\dagger(P)|0\rangle + \cdots +
\int \prod_i^5 dp_i \delta(P-\sum_i^5p_i)\psi_5(p_1,\ldots,p_5)\frac{1}{\sqrt{5!}}\prod_i^5 a^\dagger(p_i)|0\rangle
+\cdots,
\ee
in order to represent the five constituents in the intermediate states,
and we keep only the first term of $\Pfree$ and the first three terms of $\Pint$,
as the only terms that connect the two Fock sectors.
We then consider the eigenvalue problem 
\be
(\Pfree+\Pint)|\psi(P)\rangle=\left(\frac{M^2}{P}+\Pvac\right)|\psi(P)\rangle,
\ee
where we include the shift of vacuum energy $\Pvac$, to be obtained from
solving the corresponding vacuum eigenvalue problem
\be
(\Pfree+\Pint)\vac=\Pvac\vac,
\ee
with $\vac$ the lowest eigenstate.
Projection of the eigenvalue problem for the lowest massive state onto Fock sectors
yields a system of equations for the Fock-state wave functions
\bea
\label{eq:first}
\frac{\mu^2}{P}\psi_1&+&\frac{\lambda}{\sqrt{24}}\int \frac{\prod_i^4 dp_i}{4\pi\sqrt{\prod_i^4 p_i}}
\deps(\sum_i^4 p_i)\psi_5(p_1,\ldots,p_5)=\left(\frac{M^2}{P}+\Pvac\right)\psi_1, \\
\label{eq:second}
\left(\sum_i^5\frac{\mu^2}{p_i}\right)\psi_5
&+&\frac{\lambda}{24}\frac{1}{5}\left[\frac{\deps(\sum_i^4 p_i)}{4\pi\sqrt{\prod_i^4 p_i}} 
                                       + (p_5\leftrightarrow p_1,p_2,p_3,p_4)\right]\psi_1 \\
&+&20\frac{\lambda}{4}\int \frac{dp'_1 dp'_2}{4\pi\sqrt{p_1p_2p'_1p'_2}}\delta(p_1+p_2-p'_1-p'_2)\psi_5(p'_1,p'_2,p_3,p_4,p_5)
=\frac{M^2}{P}\psi_5,  \nonumber
\eea
where we have invoked a sector-dependent energy shift, with no $\Pvac$ in the top Fock 
sector.\footnote{In the top sector, there are, of course, no vacuum corrections from higher Fock sectors.}

The second equation can be solved iteratively with respect
to the self-coupling of the five-constituent Fock state
in the third term of (\ref{eq:second});
this corresponds to a diagrammatic expansion.  The leading
term in the expansion generates the vacuum bubble in Fig.~\ref{fig:tadpole}(b) that 
contributes $\Pvac$ in (\ref{eq:first}).  The second term, where the self
interaction acts once, produces the tadpole in Fig.~\ref{fig:tadpole}(a).
Both are written explicitly in (\ref{eq:bubble}) and (\ref{eq:tadpole}) below.
Subtraction of $\Pvac$ from both sides of (\ref{eq:first})
eliminates the divergent bubble.

From (\ref{eq:first}) and (\ref{eq:second}), the contributions
take the forms
\be \label{eq:bubble}
{\rm bubble}\rightarrow \int \frac{\prod_i^5 dp_i}{\prod_i^4 p_i}\delta(P-\sum_i^5p_i)
      \frac{\deps(\sum_i^4 p_i)^2}{\frac{M^2}{P}-\sum_i^5\frac{\mu^2}{p_i}}
  \sim -\int \deps(Q)^2 \frac{dQ}{\mu^2} \int \frac{\prod_i^4 dx_i}{\prod_i^4 x_i}\delta(1-\sum_i^4 x_i),
\ee
\be \label{eq:tadpole}
{\rm tadpole}\rightarrow \int \frac{\prod_i^5 dp_i \deps(\sum_i^4 p_i)}{\sqrt{\prod_i^4 p_i}}
      \frac{\delta(P-\sum_i^5p_i)}{\frac{M^2}{P}-\sum_i^5\frac{\mu^2}{p_i}}
      \int \frac{dp'_1 dp'_2}{\sqrt{p_4p_5p'_1p'_2}}
      \frac{\delta(p_4+p_5-p'_1-p'_2)}{\frac{M^2}{P}-\sum_i^3\frac{\mu^2}{p_i}-\frac{\mu^2}{p'_1}-\frac{\mu^2}{p'_2}}
      \frac{\deps(\sum_i^3 p_i+p'_1)}{\sqrt{\prod_i^3 p_i p'_1}}
\ee
The expression for the bubble diverges as $\epsilon\rightarrow0$ and is proportional
to $\delta(0)=L/4\pi$; however, the same expression is obtained for the
nontrivial vacuum energy $\Pvac$ and is subtracted.

The expression for the tadpole contribution can be simplified by
noting that $\delta(P-\sum_i^5p_i)$ reduces to $\delta(P-p_5)$,
which can be used to do the $p_5$ integral and $\delta(p_4+p_5-p'_1-p'_2)$
becomes $\delta(p_4+P-p'_1-p'_2)$, which can be used to do the $p'_2$ integral.
Finally, $\deps(\sum_i^3 p_i+p'_1)$ can be written $\deps(p_4-p'_1)$ and
used to do the $p'_1$ integral.  These leave
\be
{\rm tadpole}\sim \int\frac{\prod_i^4 dp_i }{\prod_i^4 p_i} \frac{1}{p_4P} 
   \frac{\deps(\sum_i^4 p_i)}{\left[\frac{M^2}{P}-\sum_i^4\frac{\mu^2}{p_i}-\frac{\mu^2}{P}\right]^2}
\sim \frac{1}{P}\int \deps(Q)dQ\int \frac{\prod_i^4 dx_i}{(\prod_i^4 x_i)x_4\left(\sum_i^4 \frac{\mu^2}{x_i}\right)^2},
\ee
which is finite and inversely proportional to $P$, the mark of a light-front self-energy
correction.

Of course, in a nonperturbative calculation, these contributions cannot be separated.
However, with the bubbles regulated, one can solve the eigenproblems for the vacuum
and the massive states and then carry out the necessary $\Pvac$ subtraction prior to taking
the width parameter $\epsilon$ to zero.

\section{Free scalar} \label{sec:shifted}

\subsection{Free vacuum}

The free vacuum $\vac$ is an eigenstate of the free scalar
Hamiltonian $\Pfree$ in (\ref{eq:Pminus}):
\be
\Pfree\vac=\Pvac\vac.
\ee
We will show that the vacuum is a generalized coherent state,
\be
\vac=\sqrt{Z}e^{A^\dagger}|0\rangle,
\ee
where
\be
A^\dagger=\int_0^\infty \frac{dp_1 dp_2}{\sqrt{p_1 p_2}}
     \frac{f(p_1,p_2)}{\frac{1}{p_1}+\frac{1}{p_2}}a^\dagger(p_1)a^\dagger(p_2).
\ee
For such a state, we have
\be
a(p)\vac=2\int\frac{dp'}{\sqrt{pp'}}\frac{f(p,p')}{\frac{1}{p}+\frac{1}{p'}}a^\dagger(p')\vac
\ee
and
\bea
a(p_1)a(p_2)\vac&=&\frac{2}{\sqrt{p_1 p_2}}\frac{f(p_1,p_2)}{\frac{1}{p_1}+\frac{1}{p_2}}\vac \\
&& +4\int\frac{dp'_1 dp'_2}{\sqrt{p_1 p_2 p'_1 p'_2}}\frac{f(p_1,p'_1)}{\frac{1}{p_1}+\frac{1}{p'_1}}
     \frac{f(p_2,p'_2)}{\frac{1}{p_2}+\frac{1}{p'_2}}a^\dagger(p'_1)a^\dagger(p'_2)\vac. \nonumber
\eea
With these we can apply $\Pfree$ to obtain
\bea
\Pfree\vac&=&\frac{\mu^2}{2}\int\frac{dp_1 dp_2}{\sqrt{p_1p_2}}\deps(p_1+p_2)
              \left[ a^\dagger(p_1)a^\dagger(p_2)
              +\frac{2}{\sqrt{p_1p_2}}\frac{f(p_1,p_2)}{\frac{1}{p_1}+\frac{1}{p_2}} \right. \\
 && \hspace{1in} \left.  +4\int\frac{dp'_1 dp'_2}{\sqrt{p_1 p_2 p'_1 p'_2}}
                              \frac{f(p_1,p'_1)}{\frac{1}{p_1}+\frac{1}{p'_1}}
     \frac{f(p_2,p'_2)}{\frac{1}{p_2}+\frac{1}{p'_2}}a^\dagger(p'_1)a^\dagger(p'_2)\right]\vac \nonumber \\
 &&    +\int dp \frac{\mu^2}{p}a^\dagger(p)\int dp' \frac{2}{\sqrt{pp'}}
              \frac{f(p,p')}{\frac{1}{p}+\frac{1}{p'}}a^\dagger(p')\vac.  \nonumber
\eea
The solution to $\Pfree\vac=\Pvac\vac$ is then possible if
\be
\Pvac=\frac{\mu^2}{2}\int\frac{dp_1 dp_2}{\sqrt{p_1p_2}}\deps(p_1+p_2)
       \frac{2}{\sqrt{p_1p_2}}\frac{f(p_1,p_2)}{\frac{1}{p_1}+\frac{1}{p_2}}
\ee
and the symmetrized coefficients of $a^\dagger(p_1)a^\dagger(p_2)$ sum to zero:
\bea \label{eq:coeff}
0&=&\frac{\mu^2}{2}\frac{\deps(p_1+p_2)}{\sqrt{p_1p_2}}
              +2\mu^2\int\frac{dp'_1 dp'_2}{p'_1 p'_2 \sqrt{p_1 p_2}}\deps(p'_1+p'_2)
                              \frac{f(p_1,p'_1)}{\frac{1}{p_1}+\frac{1}{p'_1}}
     \frac{f(p_2,p'_2)}{\frac{1}{p_2}+\frac{1}{p'_2}} \\
 &&    +\frac12\left[\frac{\mu^2}{p_1}+\frac{\mu^2}{p_2}\right]\frac{2}{\sqrt{p_1p_2}}
              \frac{f(p_1,p_2)}{\frac{1}{p_1}+\frac{1}{p_2}}.  \nonumber
\eea
In the second term of (\ref{eq:coeff}), we can compute
\bea
\lefteqn{\int\frac{dp'_1 dp'_2}{p'_1 p'_2}\deps(p'_1+p'_2)
                              \frac{f(p_1,p'_1)}{\frac{1}{p_1}+\frac{1}{p'_1}}
                              \frac{f(p_2,p'_2)}{\frac{1}{p_2}+\frac{1}{p'_2}}} && \\
 && =p_1 p_2 \int Q dQ \deps(Q) \int dx \frac{f(p_1,xQ)f(p_2,(1-x)Q)}{(p_1+xQ)(p_2+(1-x)Q)}
 =f(p_1)f(p_2)\int Q dQ\deps(Q)=0. \nonumber
\eea
The sum of coefficients in (\ref{eq:coeff}) is then zero if
\be
f(p_1,p_2)=-\frac12\deps(p_1+p_2).
\ee
This determines the vacuum state. 

With this solution for the coherent-state 
wave function, the energy of the vacuum is
\bea  \label{eq:Pvac}
\Pvac&=&-\frac{\mu^2}{2}\int\frac{dp_1 dp_2}{p_1p_2}
       \frac{\deps(p_1+p_2)^2}{\frac{1}{p_1}+\frac{1}{p_2}} 
     =-\frac{\mu^2}{2}\int \frac{QdQdx}{Q^2 x(1-x)}\frac{\deps(Q)^2}{\frac{1}{Q}\frac{1}{x(1-x)}} \nonumber \\
     &=&-\frac{\mu^2}{2}\int dQ\deps(Q)^2=-\frac{\mu^2}{2}\deps(0)\int_0^\infty dQ\deps(Q
         )=-\frac{\mu^2}{2}\frac{L}{4\pi}\frac12=-\frac{\mu^2 L}{16\pi}.
\eea
Here $L$ is the (infinite) volume of light-front space; however,
$\deps$ at finite $\epsilon$ regulates $\Pvac$ when it is embedded in
a nonperturbative calculation.

\begin{figure}[ht]
\vspace{0.2in}
\includegraphics[width=5cm]{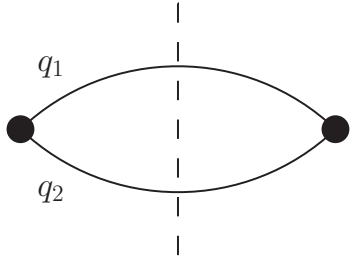}
\caption{\label{fig:oneloop}
One-loop self-energy graph as a simple vacuum bubble
which contributes to the vacuum state of a free scalar.
The dashed line indicates the intermediate state
of two ephemeral modes.
}
\end{figure}

This result is proportional to the one-loop vacuum bubble computed
by Collins~\cite{Collins}.  The equivalent perturbative calculation,
corresponding to the loop in Fig.~\ref{fig:oneloop}, is
\be
2\frac{\mu^2}{2}\int \frac{dq_1 dq_2}{\sqrt{q_1 q_2}}\deps(q_1+q_2)
\frac{1}{\frac{M^2}{P}-\frac{\mu^2}{q_1}-\frac{\mu^2}{q_2}}
\frac{\mu^2}{2}\frac{\deps(q_1+q_2)}{\sqrt{q_1 q_2}}
=2\frac{\mu^4}{4}\int \frac{dq_1 dq_2}{q_1 q_2}\frac{\deps(q_1+q_2)^2}{-\frac{\mu^2}{q_1}-\frac{\mu^2}{q_2}},
\ee
which matches (\ref{eq:Pvac}).  The leading factor of 2 comes from
the two possible contractions of the double scalar creation and
annihilation operators.

Any massive state in the free theory has a Fock-state wave function
that is a product of delta functions of the individual particle 
momenta $p_i$.  This part of the state will not mix with the 
ephemeral modes, provided the width parameter
$\epsilon$ is chosen such that it is much less than all the $p_i$.
For example, the single-particle state with mass $\mu$ and momentum
$P$ is just $a^\dagger(P)\vac$.  It is an eigenstate of $\Pfree-\Pvac$
with eigenvalue $\mu^2/P$.  Weak couplings that broaden the
momentum-space wave functions only slightly will produce effectively
unmixed contributions, but calculations with strong couplings
require more care.

\subsection{Shifted scalar}

Next we consider the shifted free scalar where $\phi\rightarrow\phi+v$.  
The new Lagrangian is
\be
{\cal L}={\cal L}_{v=0}-\mu^2 v\phi-\frac12\mu^2 v^2,
\ee
and the Hamiltonian is $\Pminus=\Pfree+\Pint$ with the interaction part
\be
\Pint=\int dx^- \mu^2 v \phi=\sqrt{4\pi}\mu^2 v \int \frac{dp}{\sqrt{p}}\deps(p)[a(p)+a^\dagger(p)]+\frac12\mu^2 v^2L.
\ee
The constant term represents the shift in the energy of the vacuum
and is therefore proportional to the spatial size $L$.

The vacuum state $\vac_v$ is now an eigenstate of $\Pminus$
\be
\left(\Pfree+\Pint\right)\vac_v=\Pvac\vac_v.
\ee
It can be constructed from the free vacuum as $e^B\vac$ with
\be
B\equiv v\int dp \sqrt{4\pi p}\,\deps(p)[a^\dagger(p)-a(p)].
\ee
This works because
\be
e^B\phi(x^-)e^{-B}=\phi(x^-)+v
\ee
and 
\be
e^B\Pfree e^{-B}=\Pfree+\Pint.
\ee
This then permits
\be
\left(\Pfree+\Pint\right)\vac_v=e^B\Pfree e^{-B}e^B\vac=e^B\Pfree\vac=\Pvac e^B\vac=\Pvac\vac_v.
\ee
Thus, in both the free and shifted cases, the vacuum is a generalized
coherent state of ephemeral modes.

The state is also correctly normalized, because
\be
_v\langle{\rm vac}\vac_v=\langle{\rm vac}|e^{B^\dagger}e^B\vac=\langle{\rm vac}|e^{-B}e^B\vac=\langle{\rm vac}\vac=1.
\ee
The vacuum expectation value of the field can also be computed:
\bea
_v\langle{\rm vac}|\phi(x^-)\vac_v&=&\langle{\rm vac}|e^{B^\dagger}\phi(x^-)e^B\vac
=\langle{\rm vac}|e^{-B}\phi(x^-)e^B\vac \nonumber \\
&=&\langle{\rm vac}|(\phi(x^-)-v)\vac=-v.
\eea
This restores the shift.

\section{Quenched scalar Yukawa theory} \label{sec:QSY}

In order to look at a case with more structure, we consider
scalar Yukawa theory~\cite{WC}, for which the Lagrangian is 
\be
{\cal L}=|\partial_\mu\chi|^2-m^2|\chi|^2+\frac12(\partial_\mu\phi)^2-\frac12\mu^2\phi^2-g\phi|\chi|^2,
\ee
where $\chi$ is a complex scalar field with mass $m$ and $\phi$ is a real scalar field
with mass $\mu$.  The two fields are coupled by a Yukawa term with strength $g$.
In two dimensions, the light-front Hamiltonian density is
\be
{\cal H}=m^2|\chi|^2+\frac12\mu^2\phi^2+g\phi|\chi|^2.
\ee
The mode expansions for the fields are (\ref{eq:phi}) for $\phi$ and
\be
\chi=\int \frac{dp}{\sqrt{4\pi p}}\left[c_+(p)e^{-ipx^-/2}+c_-^\dagger(p)e^{ipx^-/2}\right].
\ee
The nonzero commutation relations of the creation and annihilation operators are (\ref{eq:acomm}) and
\be
[c_\pm(p),c^\dagger_\pm(p')]=\delta(p-p').
\ee
In terms of these operators, the quenched light-front Hamiltonian
$\Pminus=\int dx^- {\cal H}=\Pfree+\Pint$ is specified by
\bea
\Pfree&=&\int dp \frac{m^2}{p}\left[c_+^\dagger(p)c_+(p)+c_-^\dagger(p)c_-(p)\right]
        +\int dq \frac{\mu^2}{q}a^\dagger(q)a(q) \\
  && +\frac{\mu^2}{2}\int \frac{dq_1 dq_2}{\sqrt{q_1q_2}}\deps(q_1+q_2)
                 \left[a(q_1)a(q_2)+a^\dagger(q_1)a^\dagger(q_2)\right], \nonumber
\eea
and
\be
\Pint=g\int \frac{dp dq}{\sqrt{4\pi p q (p+q)}}
  \left\{ \left[c_+^\dagger(p+q)c_+(p)+c_-^\dagger(p+q)c_-(p)\right]a(q) +{\rm h.c.}\right\}.
\ee
Pair creation and annihilation terms are suppressed for the complex scalar; without this quenching, the theory is
unstable~\cite{Baym}.  This also suppresses ephemeral modes for the complex scalar, which
would need to appear in pairs to conserve charge, leaving only those of the neutral scalar.
The vacuum in the charge-zero sector is that of the free scalar, as given in the previous
section; this provides the value of $\Pvac$ for subtraction in
the charge-one sector.  

We seek eigenstates of $\Pminus$, for which the two-dimensional light-front mass
eigenvalue problem is
\be
\Pminus|\psi(P)\rangle=\left(\frac{M^2}{P}+\Pvac\right)|\psi(P)\rangle.
\ee
We limit this to the charge-one sector.  This sector is characterized as a single 
complex scalar dressed by a cloud of neutrals.  For the present purposes we will
consider only a severe Fock-space truncation that keeps no more than two neutrals.
The Fock-state expansion for the eigenstate is then
\bea
|\psi(P)\rangle&=&\psi_0 c_+^\dagger(P)|0\rangle
   +\int dq dp \,\delta(P-q-p)\psi_1(q)a^\dagger(q)c_+^\dagger(p)|0\rangle \\
&&   +\int dq_1 dq_2 dp\, \delta(P-q_1-q_2-p)\psi_2(q_1,q_2)
          \frac{1}{\sqrt{2}}a^\dagger(q_1)a^\dagger(q_2)c_+^\dagger(p)|0\rangle. \nonumber
\eea
The normalization condition $\langle\psi(P')|\psi(P)\rangle=\delta(P'-P)$ becomes
\be \label{eq:norm}
1=|\psi_0|^2+\int dq|\psi_1|^2+\int dq_1 dq_2 |\psi_2|^2.
\ee

To construct the eigenvalue problem for the wave functions, we
act with $\Pfree$ and $\Pint$ on the eigenstate and then project onto
the three Fock sectors included in the truncation.  Terms that generate
higher Fock sectors are dropped.  For $\Pfree$ we have
\bea
\lefteqn{\Pfree|\psi(P)\rangle= \frac{m^2}{P}\psi_0c_+^\dagger(P)|0\rangle
+\int dq dp \,\delta(P-q-p)\left(\frac{\mu^2}{q}+\frac{m^2}{p}\right)\psi_1(q)a^\dagger(q)c_+^\dagger(p)|0\rangle} && \\
&&+\int dq_1 dq_2 dp \,\delta(P-q_1-q_2-p)\left(\frac{\mu^2}{q_1}+\frac{\mu^2}{q_2}+\frac{m^2}{p}\right)\psi_2(q_1,q_2)
          \frac{1}{\sqrt{2}}a^\dagger(q_1)a^\dagger(q_2)c_+^\dagger(p)|0\rangle \nonumber \\
&& +\frac{\mu^2}{2}\int\frac{dq_1dq_2}{\sqrt{q_1q_2}}\deps(q_1+q_2)\psi_0 a^\dagger(q_1)a^\dagger(q_2)c_+^\dagger(P)|0\rangle
   +\frac{\mu^2}{2}\int\frac{dq_1dq_2}{\sqrt{q_1q_2}}\deps(q_1+q_2)\psi_2(q_1,q_2)c_+^\dagger(P)|0\rangle. \nonumber
\eea
The last two terms violate momentum conservation but only by amounts of order $\epsilon$, the width of $\deps$.  
For $\Pint$ we find
\bea
\lefteqn{\Pint|\psi(P)\rangle=\frac{g}{\sqrt{4\pi}}\int\frac{dq dp}{\sqrt{qp(p+q)}}\delta(P-p-q)\left[\psi_1(q)c_+^\dagger(P)
  +\psi_0a^\dagger(q)c_+^\dagger(p)\right]|0\rangle} && \\
&&  +\sqrt{2}\frac{g}{\sqrt{4\pi}}\int\frac{dq_1 dq_2 dp}{\sqrt{q_2p(p+q_2)}}\delta(P-p-q_1-q_2)
    \left[\psi_2(q_1,q_2)a^\dagger(q_1)c_+^\dagger(p+q_2)  \right. \nonumber \\
&&  \hspace{3.3in}  \left. +\psi_1(q_1)a^\dagger(q_1)a^\dagger(q_2)c_+^\dagger(p)\right]|0\rangle.
  \nonumber
\eea
Projection of $\Pminus|\psi(P)\rangle=\left(\frac{M^2}{P}+\Pvac\right)|\psi(P)\rangle$ onto each of the
three Fock sectors yields the following three equations:
\be \label{eq:1}
\frac{m^2}{P}\psi_0+\frac{g}{\sqrt{4\pi}}\int_0^P \frac{dq\, \psi_1(q)}{\sqrt{qP(P-q)}}
+\frac{\mu^2}{\sqrt{2}}\int\frac{dq_1 dq_2}{\sqrt{q_1 q_2}}\deps(q_1+q_2)\psi_2(q_1,q_2)
=\left(\frac{M^2}{P}+\Pvac\right)\psi_0,
\ee
\be \label{eq:2}
\left(\frac{\mu^2}{q}+\frac{m^2}{P-q}\right)\psi_1(q)+\frac{g}{\sqrt{4\pi}}\frac{\psi_0}{\sqrt{qP(P-q)}}
+\sqrt{2}\frac{g}{\sqrt{4\pi}}\int_0^{P-q} \frac{dq'\,\psi_2(q,q')}{\sqrt{q'(P-q)(P-q-q')}}
=\frac{M^2}{P}\psi_1(q),
\ee
and
\bea \label{eq:3}
\lefteqn{\left(\frac{\mu^2}{q_1}+\frac{\mu^2}{q_2}+\frac{m^2}{P-q_1-q_2}\right)\psi_2(q_1,q_2)
+\frac{\mu^2}{\sqrt{2}}\deps(q_1+q_2)\frac{\psi_0}{\sqrt{q_1 q_2}}} && \\
&& +\frac{1}{\sqrt{2}}\frac{g}{\sqrt{4\pi}}\left[\frac{\psi_1(q_1)}{\sqrt{q_2(P-q_1)(P-q_1-q_2)}}
                                             +\frac{\psi_1(q_2)}{\sqrt{q_1(P-q_2)(P-q_1-q_2)}}\right]
=\frac{M^2}{P}\psi_2(q_1,q_2). \nonumber
\eea
The vacuum energy $\Pvac$ appears only in the first equation, because the Fock-space truncation
prevents any such correction in all but the lowest Fock sector.

We build a matrix representation for these equations by introducing
basis-function expansions 
\be
\psi_1(q)=\frac{1}{\sqrt{P}}\sum_n a_n f_n(q), \;\;
\psi_2(q_1,q_2)=\frac{1}{P}\sum_{nj} b_{nj} g_{nj}(q_1,q_2),
\ee
with $p=P-q$ and $p=P-q_1-q_2$, respectively, $\tilde{m}\equiv m/\mu$, and
\bea
f_{-1}(q)&=&\frac{C_{-1}}{\sqrt{q(P-q)}}\frac{P\deps(q)}{\frac{1}{q}+\frac{\tilde{m}^2}{P-q}}
    =C_{-1}\sqrt{qP}\deps(q), \\
f_n(q)&=&\frac{C_n}{\sqrt{q(P-q)}}\frac{(q/P)^n}{\frac{1}{q}+\frac{\tilde{m}^2}{P-q}},\;n\geq0,
\eea
\be
g_{-10}(q_1,q_2)=\frac{D_{-10}\sqrt{P}}{\sqrt{q_1q_2(P-q_1-q_2)}}
    \frac{P\deps(q_1=q_2)}{\frac{1}{q_1}+\frac{1}{q_2}+\frac{\tilde{m}^2}{P-q_1-q_2}}
    =D_{-10}P\sqrt{q_1q_2}\frac{\deps(q_1+q_2)}{q_1+q_2},
\ee
\be
g_{nj}(q_1,q_2)=\frac{D_{nj}\sqrt{P}}{\sqrt{q_1q_2(P-q_1-q_2)}}
    \frac{(q_1^j q_2^{(n-j)}+q_1^{(n-j)}q_2^j)/P^n}{\frac{1}{q_1}+\frac{1}{q_2}+\frac{\tilde{m}^2}{P-q_1-q_2}},\;n\geq0,j=0,\ldots,n/2.
\ee
These have the desired small-momentum behavior shown in (\ref{eq:behavior}).
The negative index $n=-1$ is reserved for the ephemeral-mode contributions.  The normalization 
condition (\ref{eq:norm}) reduces to 
\be
1=|\psi_0|^2+\sum_{nm}B_{nm}^{(1)}a_n a_m +\sum_{nj,ml}B_{nj,ml}^{(2)}b_{nj}b_{ml},
\ee
where the overlaps between the nonorthogonal basis functions are the
symmetric matrices
\be \label{eq:B1}
B_{nm}^{(1)}=\frac{1}{P}\int f_n(q) f_m(q) dq,\;\;
B_{nj,ml}^{(2)}=\frac{1}{P^2}\int g_{nj}(q_1,q_2)g_{ml}(q_1,q_2)dq_1 dq_2.
\ee
The normalization coefficients $C_n$ and $D_{nj}$ are fixed by 
requiring $B_{nn}^{(1)}=1$ and $B_{nj,nj}^{(2)}=1$, and one can
show that the $n=-1$ basis functions are orthogonal to the others,
making $B_{-1,n}^{(1)}=0$ and $B_{-10,nj}^{(2)}=0$ for $n\geq0$.

The system of equations (\ref{eq:1}-\ref{eq:3}) becomes,
with $\lambda\equiv g/\sqrt{4\pi}\mu^2$ and $\tilde{M}\equiv M/\mu$,
\be
\tilde{m}^2\psi_0-\frac{P}{\mu^2}\Pvac\psi_0+\lambda \sum_n V_n^{(0)}a_n+\sum_{nj} U_{nj}^{(0)}b_{nj}=\tilde{M}^2\psi_0,
\ee
\be
\sum_m T_{nm}^{(1)}a_m+\lambda V_n^{(0)}\psi_0+\lambda\sum_{ml}V_{n,ml}^{(1)} b_{ml}=\tilde{M}^2\sum_m B_{nm}^{(1)}a_m,
\ee
\be
\sum_{ml} T_{nj,ml}^{(2)}b_{ml}+U_{nj}^{(0)}\psi_0+\lambda\sum_m V_{m,nj}^{(1)} a_m=\tilde{M}^2\sum_m B_{nj,ml}^{(2)}b_{ml},
\ee
The various matrix elements are defined by
\be
T_{nm}^{(1)}=\int dq f_n(q)\left(\frac{1}{q}+\frac{\tilde{m}^2}{P-q}\right)f_m(q),
\ee
\be
T_{nj,ml}^{(2)}=\frac{1}{P}\int dq_1 dq_2 g_{nj}(q_1,q_2)\left(\frac{1}{q_1}+\frac{1}{q_2}+\frac{\tilde{m}^2}{P-q_1-q_2}\right)g_{ml}(q_1,q_2),
\ee
\be
U_{nj}^{(0)}=\frac{1}{\sqrt{2}}\int \frac{dq_1 dq_2}{\sqrt{q_1q_2}}\deps(q_1+q_2)g_{nj}(q_1,q_2),
\ee
\be \label{eq:V1}
V_n^{(0)}=\sqrt{P}\int\frac{dq\,f_n(q)}{\sqrt{qP(P-q}}, \;\;
V_{n,ml}^{(1)}=\sqrt{\frac{2}{P}}\int \frac{dq dq'\,f_n(q) g_{ml}(q,q')}{\sqrt{q' (P-q) (P-q-q')}}.
\ee
Details of matrix element computations are left to Appendix~\ref{sec:matrixelts}.  These include
the definition of a factor $\beta\equiv1/\left[\int_0^\infty qdq\deps(q)^2\right]$ which enters
the normalization for ephemeral modes.\footnote{The value of
$\beta$ depends on the model used for $\deps$; it is not zero because the integral is
over only half the real line.  Physical results are independent of $\beta$.}
With these matrix elements, the system of equations can be written as
\be \label{eq:psi0}
\tilde{m}^2\psi_0-\frac{P}{\mu^2}\Pvac\psi_0+\lambda\frac{\sqrt{\beta}}{2}a_{-1}
  +\lambda \sum_{n\geq0} V_n^{(0)}a_n-\frac{P\sqrt{12\beta}}{\mu^2}b_{-10}+\frac{D_{00}}{2\sqrt{2}}b_{00}
  =\tilde{M}^2\psi_0
\ee
\be  \label{eq:f-1}
n=-1:\;\;
-\frac{2P}{\mu^2}\beta\Pvac a_{-1}+\frac12\sqrt{\beta}C_0a_0+\frac12\sqrt{\beta}\lambda\psi_0
+\lambda\sqrt{12}g_{-10}=\tilde{M}^2 a_{-1}
\ee
\be
n=0:\;\;
\frac12\sqrt{\beta} C_0 a_{-1}+\sum_{m\geq0} T_{0m}^{(1)}a_m+\lambda V_0^{(0)}\psi_0
+\lambda\sum_{ml\geq0}V_{0,ml}^{(1)} b_{ml}=\tilde{M}^2\sum_{m\geq0} B_{0m}^{(1)}a_m,
\ee
\be
n>0:\;\;
\sum_{m\geq0} T_{nm}^{(1)}a_m+\lambda V_n^{(0)}\psi_0+\lambda\sum_{ml\geq0}V_{n,ml}^{(1)} b_{ml}
=\tilde{M}^2\sum_{m\geq0} B_{nm}^{(1)}a_m,
\ee
\be \label{eq:b-10}
n=-1:\;\;
-\frac{12P}{\mu^2}\beta\Pvac b_{-10}+\frac12 D_{-10}D_{00} b_{00}-\frac{\sqrt{12}P}{\mu^2}\sqrt{\beta}\Pvac\psi_0
  +\lambda\sqrt{12}a_{-1}=\tilde{M}^2 b_{-10},
\ee
\be \label{eq:b00}
n=0:\;\;
\frac12 D_{-10}D_{00}b_{-10}+\sum_{ml\geq0} T_{00,ml}^{(2)}b_{ml}
+\frac{D_{00}}{2\sqrt{2}}\psi_0+\lambda\sum_{m\geq0} V_{m,00}^{(1)} a_m=\tilde{M}^2\sum_{ml\geq0} B_{00,ml}^{(2)}b_{ml},
\ee
\be
n>0:\;\;
\sum_{ml\geq0} T_{nj,ml}^{(2)}b_{ml}+\lambda\sum_{m\geq0} V_{m,nj}^{(1)} a_m
=\tilde{M}^2\sum_{ml\geq0} B_{nj,ml}^{(2)}b_{ml}.
\ee

Cancellation of the infinite vacuum energy $\Pvac$ in (\ref{eq:psi0}), (\ref{eq:f-1}), and (\ref{eq:b-10})
is achieved if $a_{-1}=0$ and $b_{-10}=-\psi_0/\sqrt{12\beta}$.
These values correspond to the structure of the vacuum; in other words, as a part of
solving the dressed particle state, we have reconstituted
the vacuum as the foundation of the physical eigenstate and thereby
cancelled the (infinite) vacuum energy.   The projection onto $f_{-1}$, which is
equation (\ref{eq:f-1}), is no longer needed or used.  The factor
$\beta$ disappears in the eigenstate by cancelling in the 
product $b_{-10}f_{-10}\propto b_{-10}D_{-10}$ with $D_{-10}=\sqrt{6\beta}$.

This leaves a finite matrix problem with finite corrections due to nonzero
matrix elements of vacuum transitions.  In particular,
there is a finite matrix element ($D_{00}/2\sqrt{2}$) coupling the three-particle
sector ($b_{00}$) to the one-particle sector ($\psi_0$) between (\ref{eq:psi0}) 
and (\ref{eq:b00}).  The two extra particles are ephemeral modes. 

In this severe Fock-space truncation, the matrix elements are simple enough to invoke
$\epsilon\rightarrow0$ explicitly.  A more general calculation
would require a model for $\deps$ and extrapolation of
the limit $\epsilon\rightarrow0$ numerically, in
addition to consideration of several $\deps$ models
to confirm model independence.

\section{Summary}  \label{sec:summary}

We have developed a formalism by which vacuum transitions
can be included in light-front calculations and
have argued that they must be included to have
full equivalence with equal-time quantization and
to be consistent with the perturbative equivalence of
the two quantizations.  The latter equivalence
follows, at least on a formal level, as a choice
of coordinates for evaluation of Feynman diagrams,
with proper care as emphasized in~\cite{MannheimPLB,Mannheim}.  In that context, contributions
such as nonzero vacuum bubbles and tadpoles are
recovered.  These have been missing from 
nonperturbative calculations due to the neglect
of vacuum transitions in light-front Hamiltonians.
The inclusion of such transitions means that the 
light-front vacuum is not trivial and instead can be
characterized as a generalized coherent state of
ephemeral modes, even for a free theory.

Our approach is based on the realization that
vacuum transition matrix elements are nonzero with
respect to Fock-state wave functions with the correct
small-momentum behavior.  These matrix elements lead
to tadpole contributions as well as disconnected
vacuum bubbles. The vacuum bubbles are regulated by
the introduction of a finite width $\epsilon$ in
momentum-conserving delta functions, so that a
bubble's proportionality to $\delta(0)$ is 
replaced by $\deps(0)=L/4\pi$, where $L$ is the
light-front spatial volume.  The width $\epsilon$ is
taken to zero (and $L$ to infinity) after the (infinite)
vacuum energy is subtracted.  The modes with momentum
of order $\epsilon$ that are removed in this limit
are the ephemeral modes.  They represent the accumulation
of contributions at zero momentum.

The use of proper basis functions is critical.  A
standard DLCQ approximation~\cite{PauliBrodsky,BPP} cannot capture these effects,
partly because the zero-mode contributions form sets
of measure zero and partly because the DLCQ grid provides
a poor approximation to integral operators with modes
of order $\epsilon\ll P^+$, for either periodic or
antiperiodic boundary conditions.

We have considered several applications of
these ideas.  The most basic was to show
that the vacuum bubbles and tadpoles expected in $\phi^4$
theory are in fact reproduced.  We next considered the free
scalar case in detail, constructing the vacuum state as
a generalized coherent state of ephemeral modes and 
extending this to include the shifted scalar, with
recovery of the correct vacuum expectation value.
The shifted case can, of course, be handled in DLCQ by
inclusion of the constraint equation for the spatial
average of the field~\cite{Robertson,Pinsky,CH}.  Here, however,
we have an exact analytic solution with no discretization.  Also, 
the analytic solution contains the one-loop vacuum bubble
discussed by Collins~\cite{Collins} as a prime example
of light-front vacuum structure in perturbation theory.

To illustrate how the approach functions in an interacting
theory, we considered the charge-one sector of quenched
scalar Yukawa theory.  There we have shown how the vacuum
subtraction can be implemented and how strong coupling
can result in residual effects from ephemeral modes,
which in the limit translate to zero-mode effects.

This work was done in two dimensions.  The extension to
three and four dimensions should be straightforward.
The transverse momenta have the full range of $-\infty$
to $\infty$ and therefore can be balanced without being
individually zero.  The coherent state for the free scalar
vacuum would be built from an operator such as
\be
A^\dagger=\int\frac{dp_1^+ dp_2^+ d\vec{p}_\perp}{\sqrt{p_1^+p_2^+}}
   \frac{f(p_1^+,p_2^+,\vec{p}_\perp)}{\frac{1}{p_1^+}+\frac{1}{p_2^+}}
      \deps(p_1^++p_2^+)a^\dagger(p_1^+,\vec{p}_\perp)a^\dagger(p_2^+,-\vec{p}_\perp)
\ee
that creates two ephemeral modes with opposite transverse momenta.

The ideal demonstration that our approach is useful would be to
compute the critical coupling in $\phi^4$ theory.  The tadpole
contributions that were absent previously~\cite{BCH} would now be included.
Such a calculation is a natural next step.

\acknowledgments
This work was supported in part by 
the Minnesota Supercomputing Institute 
and the Research Computing and Data Services
at the University of Idaho through
grants of computing time.

\appendix

\section{Matrix elements for scalar Yukawa theory}  \label{sec:matrixelts}

We compute the matrix elements needed to resolve
the system of Fock-space equations for scalar Yukawa theory.
The matrices are defined in (\ref{eq:B1}) through (\ref{eq:V1}).
With the definition of the model-dependent\footnote{For any $\deps$ model
that scales properly with $\epsilon$, $\beta$ is independent of $\epsilon$.}
factor $\beta$
\be
\frac{1}{\beta}=\int_0^\infty dq\,q\deps(q)^2,
\ee
the basis function overlaps (\ref{eq:B1}) are, for $n,m\geq0$,
\bea
B_{-1-1}^{(1)}&=&\frac{(C_{-1})^2}{P}\int dq \, qP\deps(q)^2=\frac{(C_{-1})^2}{\beta}, \\
B_{-1n}^{(1)}&=&\frac{C_{-1}C_n}{P}\int dq\frac{\sqrt{qP}\deps(q)}{\sqrt{q(P-q)}}\frac{(q/P)^n}{\frac{1}{q}+\frac{\tilde{m}^2}{q}}
          =\frac{C_{-1}C_n}{P}\int dq\,q(q/P)^n\deps(q)\rightarrow0, \\
B_{nm}^{(1)}&=&\frac{C_nC_m}{P}\int_0^P\frac{dq}{q(P-q)}\frac{(q/P)^{n+m}}{\left(\frac{1}{q}+\frac{\tilde{m}^2}{q}\right)^2}
           =C_nC_m\int_0^1 \frac{x^{n+m+1}(1-x)dx}{(1-x+\tilde{m}^2x)^2},
\eea
\bea
B_{-10,-10}^{(2)}&=&(D_{-10})^2\int dq_1 dq_2 \, q_1 q_2\frac{\deps(q_1+q_2)^2}{(q_1+q_2)^2} 
   =(D_{-10})^2\int dx\,x(1-x)\int Q dQ\deps(Q)^2=\frac{(D_{-10})^2}{6\beta}, \nonumber \\ \\
B_{-10,nj}^{(2)}&=&\frac{D_{-10}D_{nj}}{P}\int dq_1 dq_2 \frac{\deps(q_1+q_2)}{q_1+q_2}
       \frac{q_1 q_2}{q_1+q_2}\frac{q_1^jq_2^{n-j}+q_1^{n-j}q_2^j}{P^n}\rightarrow0, \\
B_{nj,ml}^{(2)}&=&D_{nj}D_{ml} \int_0^1 dx_1 \int_0^{1-x_1} dx_2 
   \frac{(x_1^jx_2^{n-j}+x_1^{n-j}x_2^j)(x_1^lx_2^{m-l}+x_1^{m-l}x_2^j)}{x_1x_2(1-x_1-x_2)
                                          \left(\frac{1}{x_1}+\frac{1}{x_2}+\frac{\tilde{m}^2}{1-x_1-x_2}\right)^2}.
\eea
The normalization conditions, that diagonal elements of $B^{(1)}$ and $B^{(2)}$ be unity,
yield $C_{-1}=\sqrt{\beta}$ and $D_{-10}=\sqrt{6\beta}$.

The matrix elements for kinetic energy are, with $\Pvac$ defined in (\ref{eq:Pvac}),
\bea
T_{-1-1}^{(1)}&=&(C_{-1})^2\int dq\,q P\deps(q)^2\left[\frac{1}{q}+\frac{\tilde{m}^2}{q}\right]
    =(C_{-1})^2 P \int \deps(q)^2=-(C_{-1})^2\frac{2P}{\mu^2}\Pvac, \\
T_{-1n}^{(1)}&=&C_{-1}C_n\int dq\deps(q)(q/P)^n=\frac12 C_{-1}C_n\delta_{n0}, \\
T_{nm}^{(1)}&=&C_nC_m\int\frac{dx\,x^{n+m}}{1-x+\tilde{m}^2x}, \\
T_{-10,-10}^{(2)}&=&\frac{(D_{-10})^2}{P}\int dq_1 dq_2 P^2 q_1 q_2 \frac{\deps(q_1+q_2)^2}{(q_1+q_2)^2}
     \left(\frac{1}{q_1}+\frac{1}{q_2}+\frac{\tilde{m}^2}{P-q_1-q_2}\right)
     =-(D_{-10})^2\frac{2P}{\mu^2}\Pvac, \nonumber \\ \\
T_{-10,nj}^{(2)}&=&D_{-10}D_{nj}\int dq_1 dq_2\frac{\deps(q_1+q_2)}{q_1+q_2}\frac{q_1^jq_2^{n-j}+q_1^{n-j}q_2^j}{P^n}
      =\frac12 D_{-10}D_{00}\delta_{n0}, \\
T_{nj,ml}^{(2)}&=&2 D_{nj}D_{ml}\int dx_1 dx_2 \frac{x_1^jx_2^{n-j}(x_1^lx_2^{m-l}+x_1^{m-l}x_2^l)}{(x_1+x_2)(1-x_1-x_2)+\tilde{m}^2x_1x_2}.
\eea
In $T_{-1n}^{(1)}$ and $T_{-10,nj}^{(2)}$ we have used $\int dq\,\deps(q)=\frac12$, which follows
from integrating only over positive $q$.

The potential terms have the following matrix elements:
\bea
U_{-10}^{(0)}&=&\frac{D_{-10}}{\sqrt{2}}\int\frac{dq_1dq_2}{\sqrt{q_1q_2}}\deps(q_1+q_2)P\sqrt{q_1q_2}\frac{\deps(q_1+q_2)}{q_1+q_2}
    =-\frac{D_{-10}}{\sqrt{2}}\frac{2P}{\mu^2}\Pvac, \\
U_{nj}^{(0)}&=&\frac{1}{\sqrt{2}}\int\frac{dq_1dq_2}{\sqrt{q_1q_2}}\deps(q_1+q_2)
    \frac{D_{nj}\sqrt{P}}{\sqrt{q_1q_2(P-q_1-q_2)}}
    \frac{(q_1^jq_2^{n-j}+q^{n-j}q_2^j)/P^n}{\frac{1}{q_1}+\frac{1}{q_2}+\frac{\tilde{m}^2}{P-q_1-q_2}}
    =\frac{D_{00}}{2\sqrt{2}}\delta_{n0}, \nonumber \\ \\
V_{-1}^{(0)}&=&C_{-1}\sqrt{P}\int\frac{dq}{\sqrt{qP(P-q)}}\sqrt{qP}\deps(q)=\frac12 C_{-1}, \\
V_n^{(0)}&=&C_n\int\frac{dx\,x^n}{1-x+\tilde{m}^2x}, \\
V_{-1,-10}^{(1)}&=&\sqrt{2}C_{-1}D_{-10}\int dq\,q\deps(q)^2=\frac{\sqrt{2}}{\beta}C_{-1}D_{-10}=\sqrt{12}, \\
V_{n,-10}^{(1)}&=&\sqrt{2}C_nD_{-10}\int dq dq' \left(\frac{q}{P}\right)^{n+1}\frac{\deps(q+q')}{q+q'}
    =\sqrt{2}C_nD_{-10}\int dq \left(\frac{q}{P}\right)^{n+1} \deps(q)\rightarrow0, \nonumber \\  \\
V_{-1,nj}^{(1)}&=&\sqrt{2}C_{-1}D_{nj}\int \frac{dq dq'\deps(q)}{q'(P-q')}q\frac{q^j q'^{n-j}+q^{n-j}q'^j}{P^n}\rightarrow0,\\
V_{n,ml}^{(1)}&=&\sqrt{2}C_n D_{ml} \int \frac{ dx_1 dx_2 x_1^{n+1}}{1-x_1+\tilde{m}^2x_1}
        \frac{x_1^lx_2^{m-l}+x_1^{m-l}x_2^l}{(x_1+x_2)(1-x_1-x_2)+\tilde{m}^2x_1x_2}.
\eea
In $V_{n,-10}^{(1)}$ we have used a representation of the Dirac delta function
\be
\delta(q)=\int  dq' \frac{\delta(q+q')}{q+q'}
\ee
which follows from
\be
\int dq f(q) dq' \frac{\delta(q+q')}{q+q'}=\int Q dx dQ f(xQ) \frac{\delta(Q)}{Q}=\frac12 f(0).
\ee
%


\end{document}